\documentclass[11pt]{article}
\usepackage{epsfig}

\begin{document}
\pagestyle{myheadings} \markright{An alternative to Minkowski
space-time}
\title{An alternative to Minkowski space-time}
\author{Jos\'e B. Almeida\footnote{Universidade do Minho,
Departamento de F\'isica, 4710-057 Braga, Portugal. E-mail:
bda@fisica.uminho.pt}}

\maketitle

\begin{abstract}                
The starting point of this work is the principle that all
movement of particles and photons must follow geodesics of a
4-dimensional space where time intervals are always a measure of
geodesic arc lengths, i.e. $c^2 (\mathrm{d}t)^2 = g_{\alpha \beta}
\mathrm{d} x^\alpha \mathrm{d} x^\beta$, where $c$ is the speed of
light in vacuum, $t$ is time, $g_{\alpha \beta}$ is the metric
tensor and $x^\alpha$ represents any of 4 space coordinates. The
last 3 coordinates $(\alpha = 1,2,3)$ are immediately associated
with the usual physical space coordinates, while the first
coordinate $(\alpha=0)$ is later found to be related to
\emph{proper time}. Avoiding the virtually hopeless effort to
prove the initial hypothesis, the work goes through several
examples of increasing complexity, to show that it is plausible.
Starting with special relativity it is shown that there is
perfect mapping between the geodesics on Minkowski space-time and
on this alternative space. The discussion than follows through
light propagation in a refractive medium, and some cases of
gravitation, including Schwartzschild's outer metric. The last
part of the presentation is dedicated to electromagnetic
interaction and Maxwell's equations, showing that there is a
particular solution where one of the space dimensions is
eliminated and the geodesics become equivalent to light rays in
geometrical optics. A very brief discussion is made of the
implications for wave-particle duality and quantization.

\paragraph*{\hrulefill}
\begin{description}
  \item[Key words:] Alternative theories; optical propagation;
Schwartzschild; gravitation.

\end{description}

\end{abstract}

\section{Introduction}
General relativity is rooted on the consideration of Minkowski
space-time, which is adequate for the formulation of special
relativity. This space functions as tangent space in all other
situations. A consequence of this approach is that all spaces of
general relativity exhibit the characteristic $2$ signature of
Minkowski space-time and can never be reduced to Euclidean space,
which would be tangent to spaces of signature $4$. Our aim in
this paper is to suggest that special relativity situations have
an equally valid formulation in an Euclidean space, provided
appropriate coordinates are chosen. Consequently, general
situations can be studied in curved spaces having $4$ signature
and ultimately can be reduced to an Euclidean tangent space.

With the new formulation we hope to deal with equations that are
easier to solve, and generally with a simpler geometry. We will
show that there is a prospect for unification between gravity and
other interactions in nature, mainly by showing how
electromagnetic interaction can be accommodated by the theory but
also discussing some aspects related to quantum mechanics. There
is one drawback, however. The reader must not expect one-to-one
correspondence between events in relativistic space-time and
points on the space we propose, because we will not be performing
a change of coordinates, which could never change the space
signature. The physical meaning of the points of this space will
not be discussed here. The author thinks, however, that this is a
subject that must be addressed in forthcoming work, in order to
gain widespread acceptance of his proposal.

The legitimacy of the proposal stems from the following argument:
\emph{All movement of particles must follow metric geodesics of
the space when all the possible interactions are taken into
account in building up the metric. Under these circumstances two
different spaces are equally adequate for describing a particular
kind of particle movement if it is possible to map geodesics from
one space onto the other and furthermore it is possible to map
points on corresponding geodesics of both spaces} (note that this
implies a one-to-one correspondence of geodesics between both
spaces, but not a one-to-one correspondence of points). We will
establish the validity of this principle in two particular
situations of special relevance to general relativity.

First we will show that geodesics of Minkowski space-time can be
mapped onto our proposed space, thus opening the way to the use of
general curved spaces of signature $4$ for the representation of
relativistic phenomena. Later we will show that the most important
solution of general relativity for vacuum, namely Schwartzschild's
solution, also is amenable to geodesic mapping onto our proposed
space. Besides establishing the legitimacy of the proposal
through the discussion of the two mentioned cases, the paper
introduces some concepts of quantum mechanics right from the flat
space discussion, opening the road for future developments in
this area.

It is shown that optical propagation, governed by Fermat's
principle, can be expressed as propagation in a 3-dimensional
sub-space of the general 4-dimensional curved space, allowing
optical interpretations of several relativistic phenomena, namely
the interpretation of the gravitational field as a 4-dimensional
refractive index. This contributes to the validation of the
suggestion that quantum mechanical phenomena can be included in
the theory as the homologous of optical modes in waveguides.

The final sections deal with the inclusion of electromagnetic
interaction in the metric, thus allowing the expression of
charged particle trajectories as movement along geodesics. The
special case of photons is also analyzed and shown to fall
perfectly within the previously discussed situation of optical
propagation.

The whole theory is based on the hypothesis that there is always a
space where particle and photon trajectories follow metric
geodesics; this space is characterized by the universal interval
\begin{equation}
    \label{eq:interval}
    c^2 (\mathrm{d}t)^2 = g_{\alpha \beta} \mathrm{d} x^\alpha
    \mathrm{d} x^\beta,
\end{equation}
where $c$ is the speed of light in vacuum, $t$ is time, $x^0=\tau$
is a coordinate whose significance will have to be found and the
$x^a$ $(a = 1,2,3)$ correspond to the spatial cartesian
coordinates $x, y, z$. We will establish a normalization $c=1$,
which simplifies expressions and has no influence on the
generality of the discussion. Eq.\ (\ref{eq:interval})
establishes that the interval between two neighboring points of
the space divided by the time interval between the same points
equals the speed of light; this justifies the designation of
\emph{optical space} for this space.

\section{Special relativity}
The first case we consider deals with the translation of special
relativity into our new proposed framework. It is characterized
by the metric tensor
\begin{equation}
    \label{eq:srel}
    g_{\alpha \beta} = \delta_{\alpha \beta}.
\end{equation}

We want to show that, by mapping the geodesics of Minkowski
space-time to the geodesics of the optical space, there exists a
relationship between displacements on corresponding geodesics and,
therefore, any inertial movement that can be expressed in
Minkowski space-time can also be expressed in optical space. We
will use the index $\mathrm{O}$ to refer to the proposed space,
$\mathrm{O}$ standing for \emph{optical}.

Generally for any space, if the interval is expressed by
$(\mathrm{d}s)^2 = g_{\alpha \beta} \mathrm{d}x^\alpha
\mathrm{d}x^\beta$, it is possible to derive the geodesic
equation from the consideration of the Lagrangean $L = g_{\alpha
\beta} \dot{x}^\alpha \dot{x}^\beta$, with the ''dot'' indicating
derivation with respect to the arc length \cite{Inverno96}. This
Lagrangean is always constant equal to unity.

In the optical space Eq.\ (\ref{eq:interval}) establishes
$\mathrm{d}t$ as the arc length and so a geodesic can be derived
from the Lagrangean
\begin{equation}
    \label{eq:lagsrel}
    L_{\mathrm{O}} = \delta_{\alpha \beta} \dot{x}^\alpha
    \dot{x}^\beta= (\dot{\tau})^2 + (\dot{x})^2 + (\dot{y})^2 +
    (\dot{z})^2,
\end{equation}
where the "dot" is used to represent time derivatives.

The Lagrangean must be constant, equal to unity, so we can write
\begin{equation}
    \label{eq:geosrel}
    (\dot{\tau})^2 = 1 - (\dot{x})^2 - (\dot{y})^2 -
    (\dot{z})^2.
\end{equation}
This is to be compared with the equation for a geodesic in
Minkowski space-time, which can be derived from the interval
\begin{equation}
    \label{eq:mink}
    (\mathrm{d}s)^2 = (\mathrm{d}t)^2 - (\mathrm{d}{x})^2
    - (\mathrm{d}{y})^2 -   (\mathrm{d}{z})^2.
\end{equation}
For time-like geodesics $\mathrm{d}s>0$ and so, for these
geodesics, we can write
\begin{equation}
    \label{eq:lagmink}
    L_\mathrm{M} = \left( \frac{\mathrm{d}t}{\mathrm{d}s} \right)^2
    - \left( \frac{\mathrm{d}x}{\mathrm{d}s} \right)^2
    - \left( \frac{\mathrm{d}y}{\mathrm{d}s} \right)^2
    - \left( \frac{\mathrm{d}z}{\mathrm{d}s} \right)^2,
\end{equation}
where the index $\mathrm{M}$ stands for Minkowski.

Eqs.\ (\ref{eq:lagsrel}) and (\ref{eq:lagmink}) can represent the
same lines if an identification is made between homologous
quantities and $\mathrm{d}s=\mathrm{d}\tau$. In this instance
both spaces verify the relation
\begin{equation}
    \label{eq:dtaudt}
    \dot{\tau} = \sqrt{1 - (\dot{x})^2 - (\dot{y})^2 -
    (\dot{z})^2 } = \frac{1}{\gamma}~.
\end{equation}
This equation establishes a relationship between $\tau$ and
$t$,which is the same that exists between \emph{proper time} and
time for a moving particle in special relativity; we will
therefore call \emph{proper time} to the $\tau$ coordinate in our
space.

The equivalence between time-like geodesics in Minkowski
space-time and the geodesics of optical space does not go very
far in ensuring that particle movement can be studied in either
space, whenever particles are under the influence of fields that
deviate them from the 4-dimensional straight line geodesics. In
fact the results are not equivalent, but this is no limitation
because all deviation from geodesics is seen as an approximation,
which should be preferably addressed through the search for the
appropriate metric where particles will follow curved geodesics.
\section{Mass scaling of coordinates}
In the previous section we were concerned with geodesic equations
and did not consider the mass of the actual particle moving along
a geodesic. This is the usual relativistic standpoint where mass
is considered to have a role as gravitational mass, inducing
space curvature, which is separate from its role as inertial mass.
Obviously the first role does not come into play in special
relativity. We will adopt a somewhat different approach,
consistent with our initial hypothesis that time intervals always
measure arc length along the particle's trajectory.

Based on the geodesic Lagrangean equation (\ref{eq:lagsrel}), we
can define a conjugate momentum vector by
\begin{equation}
    \label{eq:geomoment}
    p_\alpha = \frac{\partial L_O}{\partial \dot{x}^\alpha}
    = 2 \delta_{\alpha \beta}  \dot{x}^\beta.
\end{equation}
The momentum as defined is not any particle's momentum but rather
a geometrical quantity related to the space geodesics. If we wish
the conjugate momentum to be related to the particle's momentum,
it is imperative to use a metric with dimensions instead of a
purely geometric one; mass must be included in the metric. This
can be achieved by a local coordinate scaling, which is required
to exhibit an extremum at the precise location of the particle.
The new metric is defined as
\begin{equation}
    \label{eq:mass}
    g_{\alpha \beta} = m^2 \delta_{\alpha \beta},
\end{equation}
with $m$ the particle's rest mass\footnote{In forthcoming
formulations it may be useful to return to a non-dimensional
metric; this is easily achieved if the metric is multiplied by
the factor $G/ ch$, with $G$ the gravitational constant and $h$
Planck's constant.}. The movement Lagrangean is conveniently
defined as $L = g_{\alpha \beta} \dot{x}^\alpha \dot{x}^\beta
/2$, allowing us to define the particle's 4-momentum as
\begin{equation}
    \label{eq:4moment}
    p_\alpha = \frac{\partial L}{\partial \dot{x}^\alpha}
    = m ^2 \delta_{\alpha \beta}  \dot{x}^\beta.
\end{equation}
The extremum condition imposed on the scale factor $m$ ensures
that it can be brought out of the derivative in Eq.\
(\ref{eq:4moment}). It will be noticed that the spatial components
of the 4-momentum correspond precisely to the classical momentum
when allowance is made for the mass scaling of the coordinates;
$p_0$ can be shown to represent the particle's kinetic energy.

There is a crucial difference between the covariant 4-momentum
defined here and the contravariant homologous from special
relativity. Our option of associating all trajectories with
geodesics leads to the association between movement and geodesic
Lagrangeans and consequently to the conjugate momentum being
defined as a covariant vector.

Mass is here defined as a coordinate scale factor which allows us
to maintain our initial hypothesis formulated in Eq.\
(\ref{eq:interval}). With this modification the particle's
4-velocity has now a magnitude equal to the speed of light
divided by the mass; the special case of zero mass will be dealt
with in section \ref{Lorentz}. It must be stressed that mass
scaling is not necessarily a purely local effect. Wherever in
space there are point masses, the local coordinates are scaled
relative to the vacuum coordinates. Nothing is said about the
effect mass has on the immediate vicinity of the traveling
particle, although we already know from general relativity that
this effect does exist. The consideration of coordinate scaling
by the traveling particle's mass will later be shown to bring
more symmetry to the equations of general relativity and
unification of the three roles of mass: Inertial mass,
gravitational active mass and gravitational passive mass.
\section{Waves in 4-dimensional optical space}
One of the main advantages of the optical space formulation of
relativity arises from the fact that it is a natural extension of
the 3-dimensional space where ray and wave optics are two
alternative ways of describing the same phenomena, wherever the
geometrical optics approach is applicable.

We feel it is useful to accompany the relativistic discussion of
particle trajectories with a discussion of wave propagation in
the same space and draw some consequences which are pertinent to
quantum mechanics. It is clear that the Euclidean 4-space of Eq.\
(\ref{eq:srel}) supports waves that verify the equation
\begin{equation}
    \label{eq:srelwave}
    \dot{\Phi}^2 =  g^{\alpha \beta}
    \partial_\alpha \partial_\beta \Phi = \delta^{\alpha \beta}
    \partial_\alpha \partial_\beta \Phi.
\end{equation}
These waves propagate in 4-space at the speed of light, which has
been normalized to unity.

If we assume $\Phi$ to be of the form $\Phi = \phi(x^\alpha) e^{
\pm i \omega t}$, Eq.\ (\ref{eq:srelwave}) becomes
\begin{equation}
    \label{eq:srelwave2}
    \left(\omega^2 \pm \delta^{\alpha \beta}
    \partial_\alpha \partial_\beta \right)
    \phi =0,
\end{equation}
and the corresponding wavelength, designated by \emph{World
wavelength}, is obviously
\begin{equation}
    \label{eq:wavelength}
    \lambda_w = \frac{2 \pi}{\omega}.
\end{equation}

We now consider the case of elementary particles to make an
identification of $\lambda_w$ in the previous equation with
Compton's wavelength $\lambda_w = h/{m c}$, with $h$ Planck's
constant; this associates a wave with every elementary particle.
Considering the speed of light normalization we have
\begin{equation}
    \label{eq:omega}
    \omega = \frac{2 \pi m }{h} = \frac{m }{\hbar}.
\end{equation}

If we temporarily remove the normalization of the speed of light,
we will quickly recognize that Eq.\ (\ref{eq:omega}) can be
rewritten as $\omega = E / \hbar$, with $E = m c^2$, which is the
same equation that links energy and frequency in the case of
photons.

\begin{figure}[thb]
    \centerline{\psfig{file=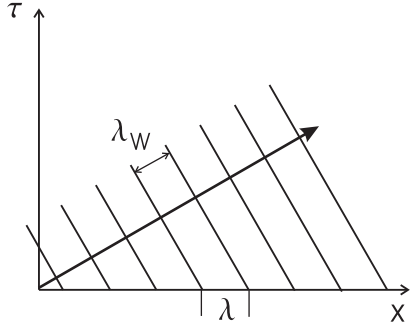, scale=0.5}}
    \caption{\label{fig:broglie}The moving particle has a world wavelength
$\lambda_w=h/(mc)$ and a spatial wavelength $\lambda=h/(mv)$.}
\end{figure}

Fig.\ \ref{fig:broglie} illustrates a particle's trajectory with
superimposed wavefronts spaced $\lambda_w$ apart; these wavefronts
intersect 3-dimensional physical space defining a wavelength
$\lambda$ which, considering Eq.\ (\ref{eq:dtaudt}) is
\begin{equation}
    \label{eq:broglie}
    \lambda = \frac{\lambda_w }{\sqrt{(\dot{x})^2 (\dot{y})^2
    (\dot{z})^2}}=\frac{h}{p}~;
\end{equation}
$p$ represents the magnitude of the spatial component of momentum
and this equation is exactly the definition of De Broglie's
wavelength. We have thus established a relation between Compton's
and De Broglie's wavelength as a purely geometric one.

In the previous derivation we have assumed that particles move
through 4-space with a 4-velocity with the magnitude of the speed
of light and have an associated wavelength given by Eq.\
(\ref{eq:wavelength}). If we now allow mass scaling of the
coordinates as implied by Eq.\ (\ref{eq:mass}), Eq.\
(\ref{eq:srelwave}) must be replaced by
\begin{equation}
    \label{eq:srelwave3}
    \dot{\Phi}^2 = g^{\alpha \beta}
     \partial_\alpha \partial_\beta
    \Phi = \frac{1}{m^2}
    \delta^{\alpha \beta} \partial_\alpha \partial_\beta
    \Phi.
\end{equation}
The wavelength becomes independent of the particle's mass and
universally given by Planck's constant.

\section{\label{Lorentz}Lorentz equivalent transformations}
Our approach to coordinate transformation between observers moving
relative to each other is different from special relativity.
While in the latter case the interval is given by Eq.\
(\ref{eq:mink}), thus ensuring that a coordinate transformation
preserves the interval and affects both spatial and time
coordinates, our option of making time intervals measure geodesic
arc length gives time a meaning independent of any coordinate
transformation. We thus propose that Lorentz equivalent
transformations between a "fixed" or "laboratory" frame and a
moving frame are performed in two separate phases. The first
phase is a tensorial coordinate transformation, which changes the
coordinates keeping the origin fixed, with no influence in the way
time is measured, while the second phase corresponds to a "jump"
into the moving frame, changing the metric but not the
coordinates.

Making use the metric from Eq.\ (\ref{eq:mass}) it is possible to
express $\dot{x}^0$ in terms of $\dot{x}^a$, in a similar way to
what was used to write Eq.\ (\ref{eq:geosrel}):
\begin{equation}
    \label{eq:geosrel2}
    m^2 \left(\dot{x}^0\right)^2 = 1 - m^2 \delta_{a b}
    \dot{x}^a \dot{x}^b.
\end{equation}
What is immediately obvious is that particles with zero mass,
such as photons, cannot follow a geodesic of this space unless,
as a limiting case, we force them to follow geodesics with
$\mathrm{d} x^0 = 0$. We will deal with optical propagation in
the next section; for now it will be sufficient to know that
photons travel on the 3-dimensional $x^a$ space. Photons carry the
value of the $x^0$ coordinate and can be used to
\emph{synchronize} all points in space to the observer's own
measurements of this coordinate. Following this argument we can
say that a coordinate transformation must preserve the value of
$\mathrm{d} x^0$. It will readily be seen that this statement is a
direct equivalent to interval invariance in special relativity.
On the other hand the time interval must also evaluate to the
same value on all coordinate systems of the same frame, due to its
definition as interval of the optical space.

Let us consider two unit mass observers $O$ and $\bar{O}$, the
latter moving along one geodesic of $O$'s coordinate system. Let
the geodesic equation have a parametric equation $x^{\alpha'}
(t)$. Our aim is to establish the coordinate transformation tensor
between the two observers' coordinate systems,
${\Lambda^{\bar{\mu}}}_\nu = \partial x^{\bar{\mu}} / \partial
x^\nu$; we have already established that $\mathrm{d} x^{\bar{0}}
= \mathrm{d} x^0$ and so it must be ${\Lambda^{\bar{0}}}_0 = 1$
and ${\Lambda^{\bar{0}}}_a = 0$.

A photon traveling parallel to the $x^1$ direction follows a
geodesic characterized by $\mathrm{d}t = \mathrm{d} x^1$ on $O$'s
coordinates. On $\bar{O}$'s coordinates the photon will move
parallel to $x^{\bar{1}}$ and so we can say that it is also
$\mathrm{d}t = \mathrm{d} x^{\bar{1}}$. The same behaviour could
be established for all three $(x^a,~x^{\bar{a}})$ coordinate pairs
and we conclude that ${\Lambda^{\bar{a}}}_a = 1$.

\begin{figure}[thb]
    \centerline{\psfig{file=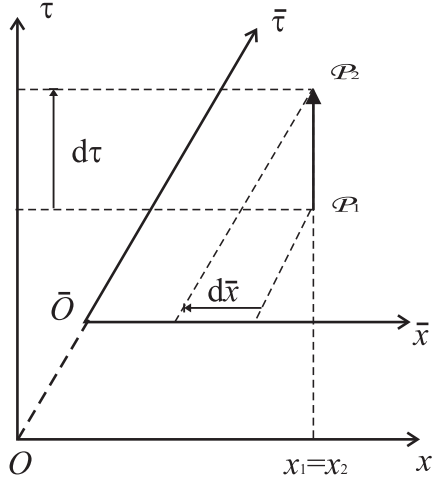, scale=0.7}}
    \caption{\label{fig:Lorentz}The worldline of moving observer $\bar{O}$ coincides with the
    $\bar{\tau}$ axis, while the $\bar{x}$ axis stays parallel to the $x$ axis.
    A displacement parallel to axis $\tau$ implies both $\bar{x}$ and $\bar{\tau}$
    components in the moving frame.}
\end{figure}

Consider now two points on a line parallel to $x^0$ axis in $O$'s
coordinates, separated by a time interval $\mathrm{d}t =
\mathrm{d} x^0$. On $\bar{O}$'s coordinates it will be
\begin{equation}
    \mathrm{d} x^{\bar{a}} = -\dot{x}^{a'} \mathrm{d}t
    =\frac{\dot{x}^{a'}}{\dot{x}^{0'}}~\mathrm{d} x^0,
\end{equation}
allowing us to conclude that ${\Lambda^{\bar{a}}}_0 =
-\dot{x}^{a'}/\dot{x}^{0'}$. Fig.\ \ref{fig:Lorentz} shows
graphically the relation between the two coordinate systems.

If we now allow the two observers to have different mass and
introduce  mass scaling on the coordinate transformation we define
the coordinate transformation tensor between $x^\nu$ and
$x^{\bar{\mu}}$ as
\begin{equation}
    \label{eq:transform}
    {\Lambda^{\bar{\mu}}}_\nu = \frac{\bar{m}}{m}\left[\begin{array}{cccc}
      1 & 0 & 0 & 0 \\
      -\dot{x}^{1'}/\dot{x}^{0'} & 1 & 0 & 0 \\
      -\dot{x}^{2'}/\dot{x}^{0'} & 0 & 1 & 0 \\
      -\dot{x}^{3'}/\dot{x}^{0'} & 0 & 0 & 1 \
    \end{array} \right],
\end{equation}
where $m$ and $\bar{m}$ are the masses of observers $O$ and
$\bar{O}$, respectively.

We are now in position to evaluate the metric for $\bar{O}$'s
coordinates
\begin{equation}
    \label{eq:movmetr}
    g_{\bar{\mu} \bar{\nu}} = {\Lambda^{\bar{\mu}}}_\alpha
    {\Lambda^{\bar{\nu}}}_\beta g_{\alpha \beta};
\end{equation}
after evaluation we get
\begin{equation}
    \label{eq:movmetr2}
    g_{\bar{\mu} \bar{\nu}} = \left(\frac{\bar{m}}{m}\right)^2
    \left[\begin{array}{cccc}
      1 + \delta_{a' b'}\breve{x}^{a'}\breve{x}^{b'}
      & -\breve{x}^{1'}
      & -\breve{x}^{2'} & -\breve{x}^{3'} \\
      -\breve{x}^{1'} & 1 & 0 & 0 \\
      -\breve{x}^{2'} & 0 & 1 & 0 \\
      -\breve{x}^{3'} & 0 & 0 & 1 \
    \end{array}\right],
\end{equation}
with $\breve{x}^{a'}=\dot{x}^{a'}/\dot{x}^{0'}$.

The metric given by the previous equation will evaluate time in
$\bar{O}$'s coordinates in a frame which is fixed relative to
$O$, that is, this is still time as measured by observer $O$.
Observer $\bar{O}$ will measure time intervals
$\mathrm{d}\bar{t}$ in his own frame and so, although the
coordinates are the same as they were in $O$'s frame, time is
evaluated with the standard metric $\bar{m}^2 \delta_{\bar{\mu}
\bar{\nu}}$ instead of metric given by Eq.\ (\ref{eq:movmetr2}).

Mathematically the metric replacement corresponds to the identity
\begin{equation}
    \label{eq:identity}
    \delta_{\bar{\alpha} \bar{\beta}} =
    {\Lambda_{\bar{\alpha}}}^{\bar{\mu}}
    {\Lambda_{\bar{\beta}}}^{\bar{\nu}} g_{\bar{\nu}\bar{\beta}};
\end{equation}
this operation will be used later to derive the metric for
electromagnetic interaction. We note that this transformation
preserves the essential relationship of special relativity, i.\
e. for a unit mass particle $\mathrm{d}\bar{t}^2 - \delta_{\bar{a}
\bar{b}}\mathrm{d}x^{\bar{a}}\mathrm{d}x^{\bar{b}} =
\mathrm{d}t^2-\delta_{a b} \mathrm{d} x^a \mathrm{d} x^b$.

\section{Optical propagation}
Let us now consider the situation where the metric has the form
\begin{equation}
    \label{eq:optical}
    g_{0\alpha}=g_{\alpha 0}=0,~~ g_{a b} = n^2 \delta_{a
    b}~ (a, b \neq 0),
\end{equation}
with $n$ a function of the coordinates. This is really the metric
of a 3-dimensional subspace where the coordinate $x^0$ does not
intervene. Eq.\ (\ref{eq:interval}) becomes
\begin{equation}
    \label{eq:Fermat}
    (\mathrm{d}t)^2 = n^2\left[(\mathrm{d}x)^2 + (\mathrm{d}y)^2 +
    (\mathrm{d}z)^2\right],
\end{equation}
which is readily seen to lead to Fermat's principle when this
variational principle is written
\begin{equation}
    \label{eq:Fermat2}
    \delta \int n \mathrm{d} \sigma  = 0,
\end{equation}
with $n$ the refractive index of the optical medium and
$\mathrm{d}\sigma$ the interval in Euclidean space given by
\begin{equation}
    \mathrm{d}\sigma = \sqrt{(\mathrm{d}x)^2 + (\mathrm{d}y)^2
    + (\mathrm{d}z)^2}.
\end{equation}

Obviously this space supports 3-dimensional waves, as it is none
other than the usual optical space where all the rules of optical
propagation are well established.

The propagation speed is $1/n$ and the corresponding wavelength is
$\lambda = h / n E$, with $E$ the photon energy.
\section{Gravitation}
Stationary vacuum solutions for gravitational field are expected
to have a null Ricci tensor, meaning that a suitable coordinate
change will produce a diagonal metric with just $1$ and $-1$
elements in the diagonal \cite{Inverno96}; this combined with the
characteristic $4$ signature of the optical space we are using
means that these solutions must be conformal transformations of
an Euclidean metric. We define these solutions by the general
metric
\begin{equation}
    \label{eq:gravit}
     g_{\alpha \beta} = n^2 \delta_{\alpha \beta},
\end{equation}
where $n$ is a function of the coordinates. This is a 4D extension
of the optical 3D sub-space discussed in the previous section and
further justifies the choice of \emph{optical space} as the
designation for our proposed 4-space.

Similarly to what was done for the Minkowski space-time, it is
possible to map the geodesics of this particular situation to
time-like geodesics of general relativity's curved space-time for
vacuum. Eq.\ (\ref{eq:interval}) now becomes
\begin{equation}
    \label{eq:intgravit}
    (\mathrm{d}t)^2 = n^2 \left(\mathrm{d}\tau^2 + \mathrm{d}x^2
    + \mathrm{d}y^2+ \mathrm{d}z^2\right);
\end{equation}
which can be arranged as
\begin{equation}
    \label{eq:grel}
    (\mathrm{d}s)^2 = \frac{1}{n}~ (\mathrm{d}t)^2-n \left(\mathrm{d}x^2
    + \mathrm{d}y^2+ \mathrm{d}z^2\right).
\end{equation}

The equation above can be identified with the metric definition
for several situations of gravitational space curvature. For
instance Schwartzschild's outer metric will be obtained by first
finding metric of the Eq.\ (\ref{eq:gravit}) type which is
adequate for Newtonian mechanics, and imposing the solution to be
spherically symmetric and asymptotically Euclidean for large
ones. The solution is eventually
\begin{equation}
    \label{eq:schwart}
    \frac{1}{n} = \left(1 - \frac{ G M}{ r } \right).
\end{equation}
Here $G$ is the gravitational constant, $M$ is the mass of a
large body and $r$ is the distance to the body\footnote{The
angular coordinates of this solution are not the same as
Schwartzschild's but can be converted by a simple coordinate
transformation.}.

Recalling Eq.\ (\ref{eq:grel}) and making $(\mathrm{d}s)^2 = n
(\mathrm{d}\tau)^2$, we can invoke the Lagrangean unity to arrive
at the relation
\begin{equation}
    \label{eq:dtaudtgravit}
    \dot{\tau} = \sqrt{\frac{1}{n^2} - \dot{x}^2 - \dot{y}^2
    - \dot{z}^2} = \sqrt{\frac{1}{n^2} - v^2}.
\end{equation}

Eq.\ (\ref{eq:gravit}) is obtained from Eq.\ (\ref{eq:srel}) by
multiplication of the right-hand side by $n^2$, equivalent to a
position dependent scale factor. When $n$ is a constant
independent of the coordinates the space becomes flat and only the
relationship between time and space intervals is altered; this is
the same effect that was produced by the particle's own mass in
the special relativity discussion. We are then led to conclude
that mass must be a source of space curvature leading eventually
to Einstein's equations modified for the optical space. This will
be the subject of a future publication and will not be pursued in
the present paper.

Naturally it is possible to extend Fermat's principle to this
4-dimensional space, deriving particle's trajectories as if they
were 4-dimensional light rays. Just as the refractive index does
in optics, gravity can be interpreted as slowing down the
4-dimensional speed of the particles relative to its $c/m$ value
away from other masses. Similarly to optics, the refractive index
approach is an alternative to space curvature.

The parallel with optics becomes especially interesting when we
associate to every elementary particle a beat frequency $f=m
c^2/h$, as suggested before. The particle's worldline is then
marked by a wavelength known as Compton wavelength and 3D
projection of this wavelength is found to define the De Broglie
wavelength, as we saw earlier. In situations where particles are
confined to short radii or the gravitational field is high, thus
lengthening their Compton or World wavelength, 4D wave modes start
to become important and lead ultimately to quantization.
\section{Electrostatic field}
Let us now turn our attention to electrostatic field by
consideration of the electrostatic force on a charged particle of
mass $m$ and electric charge $q$. This force can be written as
\begin{equation}
    \label{eq:elforce}
    \vec{f} = -q \nabla V.
\end{equation}
Here $V$ is the electrostatic potential such that $\nabla V =
\vec{E}$ with $\vec{E}$ the electric field. The arrows were used
to denote vectors in 3D space and the ''nabla'' operator has its
usual 3-dimensional meaning. If the particle is under the single
influence of the electrostatic force, the rate of change of its
momentum will equal this force and so we write $m~
\mathrm{d}^2\vec{r}/\mathrm{d} t^2 = -q \nabla V$; $\vec{r}$ is
the position vector. If we use the coordinate scaling by mass
implied by Eq.\ (\ref{eq:mass}), the spatial components of the
momentum vector must appear as $m^2 \delta_{a b} \ddot{x}^b$; the
gradient is also affected by the scaling and we expect it to
appear as $\partial_a V /m$. If we use primed indices to denote
unscaled coordinates it is
\begin{equation}
    m \delta_{a' b'}\ddot{x}^{b'} = -q \partial_{a'} V,
\end{equation}
\begin{equation}
    \label{eq:elmomentum}
    m^2 \delta_{a b}\ddot{x}^b = -\frac{q}{m}~\partial_a V.
\end{equation}

We are looking for a Lagrangean of the form
\begin{equation}
    \label{eq:elproposed}
    L = g_{00} \left(\dot{x}^0\right)^2 + m^2 \delta_{a b} \dot{x}^a
    \dot{x}^b,
\end{equation}
which is consistent with the non-relativistic form of the
electrostatic force. For speeds much smaller than the speed of
light $(\dot{x}^0)^2 \simeq 1/g_{00}$. If we derive the
Euler-Lagrange's equations for the 3 spatial dimensions we get
\begin{equation}
    \label{eq:eleuler}
   2 m^2 \delta_{a b}\ddot{x}^b = \partial_a
    g_{00} \left(\dot{x}^0\right)^2,
\end{equation}
and the compatibility between Eqs.\ (\ref{eq:elmomentum} and
\ref{eq:eleuler}) suggests that we make $g_{00}= m^2
\mathrm{exp}(-2qV/m)$.

The metric for the electrostatic situation follows directly
\begin{equation}
    \label{eq:elmetric}
      g_{\alpha \beta} = m^2 \left[\begin{array}{cccc}
      V_0 & 0 & 0 & 0 \\
      0 & 1 & 0 & 0 \\
      0 & 0 & 1 & 0 \\
      0 & 0 & 0 & 1 \
    \end{array}\right],
\end{equation}
with $V_0 = \mathrm{exp}(-2qV/m)$.

The geodesics of the space defined above correctly predict
particle movement under an electrostatic field in the
non-relativistic situation and provide a plausible generalization
for the relativistic cases. The relativistic prediction is not
entirely coincident with those based on general relativity,
allowing deviation from geodesics. The proposed optical space's
predictions are equivalent to general relativity only in those
cases when movements follow geodesics; the allowance that is made
in general relativity for parallel transport can also be made
here as an approximation to the desired approach, which involves
finding the metric for each and every particle movement. When
particles are allowed to deviate from geodesics we do not expect
a perfect match between results obtained in the optical and
relativistic spaces.

When the electric potential exhibits a minimum restricted to a
small region of space, the particle can enter a closed orbit; if
we recall the 4D optical analogy, the orbit can be seen as
analogous to a light ray being confined to an optical waveguide.
It is no wonder that propagation modes start to develop as the
orbit diameter decreases, which are the 4D counterparts of
quantum phenomena.

The wave equation for an elementary particle can be written
\begin{equation}
    \label{eq:electwave}
    \dot{\Phi}^2 = g^{\alpha \beta}
     \partial_\alpha \partial_\beta
    \Phi = \frac{1}{m^2}\left[\frac{\left(\partial_0 \Phi
    \right)^2}{V_0}+
    \delta^{a b} \partial_a \partial_b
    \Phi\right].
\end{equation}
If the partial derivative $\partial_0 \Phi$  is expressed in
terms of $\dot{\Phi}$ and $\partial_a \Phi$ through the metric,
the equation will become a 3-dimensional wave equation which has
been shown to degenerate in Schr\"odinger equation in the
non-relativistic limit \cite{Almeida00:4}. Outside this limit Eq.\
(\ref{eq:electwave}) will produce a relativistic 3D wave equation
which is expected to be compatible with quantization due to
electrostatic force.
\section{Electromagnetism and light propagation}
In a non-relativistic situation the Lorentz force on a moving
particle of electric charge $q$ can be written $\vec{f} = q
(\vec{E} + \vec{v} \times \vec{B})$, where the arrows were used to
identify vectors in 3D non-relativistic mechanics, $\vec{E}$ is
the electric field, $\vec{B}$ the magnetic field and $\vec{v}$ the
particle's speed. We now use Einstein's argument
\nocite{Lorentz83} \cite{Einstein05} to say that if we place our
frame on the moving particle this will be under the influence of
an electrostatic force, due to the zeroing of the velocity on the
Lorentz force expression. The electromagnetic force should then
be the consequence of expressing the electrostatic force on a
frame that is not moving with the particle.

We then use Eqs.\ (\ref{eq:movmetr} and \ref{eq:identity}) to
transform the metric of Eq.\ (\ref{eq:elmetric}) into the
electromagnetic situation; the bar over the indices indicates the
particle frame.
\begin{equation}
    \label{eq:elmagmetric}
    g_{\alpha \beta} = {\Lambda_\alpha}^{\bar{\mu}}
    {\Lambda_\beta}^{\bar{\nu}} g_{\bar{\mu} \bar{\nu}}
    = m^2\left[\begin{array}{cccc}
      V_0 & V_1 & V_2 & V_3 \\
      V_1 & 1 + \frac{(V_1)^2}{V_0} &  \frac{V_1 V_2}{V_0} &  \frac{V_1 V_3}{V_0} \\
      V_2 &  \frac{V_1 V_2}{V_0} & 1 + \frac{(V_2)^2}{V_0} &  \frac{V_2 V_3}{V_0} \\
      V_3 &  \frac{V_1 V_3}{V_0} & \frac{V_2 V_3}{V_0} & 1 + \frac{(V_3)^2}{V_0} \
    \end{array} \right],
\end{equation}
with
\begin{equation}
    \label{eq:vetorpot}
    V_0 = e^{-2qV/m}~, ~~~~ V_a = V_0 \frac{\dot{x}^{a'}}{\dot{x}^{0'}}.
\end{equation}
Note that the $\dot{x}^{\alpha '}$ of the previous equation refer
to the movement of the frame where the Lorentz force becomes
purely electrostatic and not to any particle's movement.

We can now write the Lagrangean for a geodesic of the
electromagnetic space using Eq.\ (\ref{eq:elmagmetric}), the
$x^\alpha$ representing the inertial movement of a particle under
electromagnetic force:
\begin{equation}
    \label{eq:lagmag}
    L = g_{\alpha \beta} \dot{x}^\alpha \dot{x}^\beta.
\end{equation}

The electromagnetic field can be associated to an anti-symmetric
tensor $F_{\alpha \beta}$ such that \cite{Inverno96}
\begin{equation}
    \label{eq:ftensor}
    F_{\alpha \beta} = \left[\begin{array}{cccc}
      0 & -E_1 & -E_2 & -E_3 \\
      E_1 & 0 & B_3 & -B_2 \\
      E_2 & -B_3 & 0 & B_1 \\
      E_3 & B_2 & -B_1 & 0 \
    \end{array}\right].
\end{equation}
$F_{\alpha \beta}$ can be obtained from $V_\alpha$ through
\begin{equation}
    \label{eq:ftensor2}
    F_{\alpha \beta} = \frac{m}{2q V_0}\left(\partial_\alpha V_\beta -
    \partial_\beta V_\alpha \right).
\end{equation}

If we start with the field tensor referred to a frame where the
Lorentz force becomes purely electrostatic and use the
transformation tensor to refer to the fixed frame we can write
\begin{equation}
    \label{eq:ftensorfixed}
    F_{\alpha \beta} = {\Lambda_\alpha}^{\bar{\mu}}
    {\Lambda_\beta}^{\bar{\nu}}F_{\bar{\mu} \bar{\nu}},
\end{equation}
with
\begin{equation}
    \label{eq:ftensormob}
    F_{\bar{\mu} \bar{\nu}}=  \left[\begin{array}{cccc}
      0 & -E_1 & -E_2 & -E_3 \\
      E_1 & 0 & 0 & 0 \\
      E_2 & 0 & 0 & 0 \\
      E_3 & 0 & 0 & 0 \
    \end{array}\right];
\end{equation}
the final result is
\begin{equation}
    \label{eq:ftensorfixed2}
    F_{\alpha \beta} = \left[\begin{array}{cccc}
      0 & -E_1 & -E_2 & -E_3 \\
      E_1 & 0 & E_2 \breve{x}^1 - E_1 \breve{x}^2  &
       E_3 \breve{x}^1 - E_1 \breve{x}^3 \\
      E_2 & E_1 \breve{x}^2 - E_2 \breve{x}^1 &
      0 & E_3 \breve{x}^2 - E_2 \breve{x}^3 \\
      E_3 & E_1 \breve{x}^3 - E_3 \breve{x}^1 &
      E_2 \breve{x}^3 - E_3 \breve{x}^2 & 0 \
    \end{array}\right].
\end{equation}

Let us look at a particular solution pertaining to the propagation
of electromagnetic radiation, which should have a null component
of momentum along the $x^0$ direction. Considering Eq.\
(\ref{eq:lagmag}) the first component of the momentum vector is
\begin{equation}
    \label{eq:p0mag}
    p_0 = 2 m^2  V_\alpha \dot{x}^\alpha
    = 2 m^2\left(V_0 \dot{x}^0 + V_a \dot{x}^a\right).
\end{equation}
The particular solution we are searching calls for $p_0=0$
\begin{equation}
    \label{eq:p0part}
    \dot{x}^0=- \frac{V_a \dot{x}^a}{V_0 }.
\end{equation}
We note immediately that if $x^0$ is not to grow indefinitely it
must average zero and so we postulate that it is a periodic
function of $t$. Eq.\ (\ref{eq:p0part}) can be replaced in Eq.\
(\ref{eq:lagmag})
\begin{equation}
    \label{eq:lagpart}
    L = m^2 \delta_{a b} \dot{x}^a \dot{x}^b .
\end{equation}

The result is a Lagrangean which depends only on the spatial
variables and is a special case of the optical propagation
condition given by Eq.\ (\ref{eq:optical}); here $m^2$ plays the
role of the refractive index. In fact an optical medium is
expected to have a complex metric and the final refractive index
will be the result of all the contributions, including the mass
distribution. It is noticeable that if we had derived Eq.\
(\ref{eq:lagpart}) in a gravitational field situation, this would
appear as an $n^2$ factor in the second member, accounting for
the redshift induced by gravity and light bending near massive
bodies.
\section{Conclusions and further developments}
We proposed two main premises for relativistic mechanics, these
being: \emph{''All trajectories will follow geodesics in a
suitably defined 4-dimensional space''} and \emph{''Geodesic or
trajectory length can be measured by the time interval multiplied
by the speed of light in vacuum.''} We find it virtually
impossible to demonstrate the validity of those premises and so
we chose to show that they will produce the same consequences as
General relativity in two particularly important situations,
namely Minkowski and Scwartzchild metrics. The equivalence to
General relativity was based on the argument that \emph{''If
particle's trajectories follow metric geodesics, the mapping of
geodesics between two spaces is a sufficient condition for those
spaces to be equivalent from the point of view of particle
movement.''}

We also showed that optical propagation follows the same rules as
particle trajectories but is restricted to a 3-dimensional
sub-space. The association of a 4-dimensional wave equation to
the trajectories of elementary particles, in a similar way to the
3-dimensional waves associated with photons, allowed the
derivation of an important connection between Compton's and De
Broglie's wavelengths as a purely geometrical one. Quantization
was also shown to result whenever particles are restricted to
small orbits or to orbits under strong fields.

The Lorentz force and electromagnetism were also shown to be
compatible with the initial premises, allowing the prediction of
trajectories under electromagnetic interaction and the connection
between light propagation in optical media, the metric and
optical refractive index.

Two main directions of forthcoming work are expected to produce
results and contribute to validate the theory. One area of work
will try to establish equations equivalent to Einstein's in this
new formulation. This will be a direct consequence of the
coordinate scaling by the mass that was introduced in this paper
for point particles. A straightforward generalization will
replace mass by mass density and coordinate scaling by curvature.
It is expected that this line of work will yield further
validation but most of all it is expected to lead to equations
that are easier to solve then Einstein's in a variety of
situations.

A different aspect will be the exploitation of the 4-dimensional
wave equation, namely for particle interactions. We feel that we
have only skimmed the surface of this rich field and that there
is scope for a large number of important results integrating
gravitation with electromagnetism and eventually with all the
known particle interactions.

  \bibliography{aberrations}   

\begin{thebibliography}{1}

\bibitem{Inverno96}
Ray D'{I}nverno.
\newblock {\em Introducing Einstein's Relativity}.
\newblock Clarendon Press, Oxford, 1996.

\bibitem{Almeida00:4}
Jos{\'e}~B. Almeida.
\newblock Optical interpretation of special relativity and quantum mechanics.
\newblock In {\em {OSA} Annual Meeting}, Providence, RI, 2000.
\newblock http://www.arXiv.org/abs/physics/0010076.

\bibitem{Lorentz83}
H.~A. Lorentz, A.~Einstein, and Minkowski, editors.
\newblock {\em The Principle of Relativity}, volume~1 of {\em Textos
  Fundamentais da F\'isica Moderna}, Lisboa, 1983. Funda\c{c}{\~a}o Calouste
  Gulbenkian.
\newblock \emph{O Princ\'ipio da Relatividade}, Portuguese translation.

\bibitem{Einstein05}
A.~Einstein.
\newblock About the electrodynamics of moving bodies.
\newblock In Lorentz et~al. \cite{Lorentz83}, pages 47--86.
\newblock Portuguese translation of Ann. d. Phys. \textbf{17} (1905).

\end{thebibliography}
  \bibliographystyle{unsrt}

\end{document}